\documentclass[11pt]{revtex4}
\usepackage{amsmath}
\usepackage{amsfonts}
\usepackage{amssymb}
\begin{document}
\title{Degrees of freedom of massless boson and fermion fields in any even dimension~%
\footnote{This is the talk published in 
the Proceedings to the $18^{th}$ Workshop "What Comes Beyond the Standard Models",
Bled, 11-19 of July, 2015.}}
\author{N.S. Manko\v c Bor\v stnik${}^1$ and H.B.F. Nielsen${}^2$\\
${}^1$University of Ljubljana,\\
Slovenia%\\
${}^2$Niels Bohr Institute,\\
Denmark
%Copenhagen, DK-2100
}
% To be changed
\begin{abstract}
This is a discussion on degrees of freedom of massless fermion and boson fields, if they are free or 
weakly interacting.  We generalize the gauge fields of $S^{ab}$ - $\omega_{abc}$ - and of 
$\tilde{S}^{ab}$ - $ \tilde{\omega}_{abc}$ - of the {\it spin-charge-family} to the gauge fields 
of all possible products of $\gamma^a$'s and of all possible products of $\tilde{\gamma}^a$'s,
the first taking care in the {\it spin-charge-family} theory of the spins and charges quantum numbers
($\tau^{Ai}=\sum_{a,b} c^{Ai}{}_{ab} \,S^{ab}$) of fermions, the second ($\tilde{\tau}^{Ai}=
\sum_{a,b} \tilde{c}^{Ai}{}_{ab}\, \tilde{S}^{ab}$) 
taking care of the families quantum numbers. 
\end{abstract}
%\maketitle
%



\maketitle

\section{Introduction}
\label{introduction} 

The purpose of this contribution to the Discussion section of this Proceedings to the Bled 2015
 workshop is to understand, hopefully  (at least a little) better: 
{\bf a.}  Why is the simple starting action of the {\it spin-charge-family} theory%
~\cite{n2014matterantimatter,n2012scalars,JMP,scalarfields2015JMP} doing so well in 
manifesting the observed properties of the fermion and boson fields? {\bf b.} Under which condition 
would more general action lead to the starting action of Eq.~(\ref{wholeaction})?  {\bf c.} What 
would more  general action, if leading to the same low energy physics, mean for the history of our 
Universe? {\bf d.} Could the fermionization procedure of boson fields or the bosonization procedure 
of fermion fields, discussed in this Proceedings for any dimension $d$ (by the authors of this 
contribution, while one of them, H.B.F.N.~\cite{holger},  has succeeded with another author to do
the fermionization for $d=(1+1)$), help to find the answers to the questions under  {\bf a. b. c.}?

In the {\it spin-charge-family} theory of one of us (N.S.M.B.)~\cite{n2014matterantimatter,%
n2012scalars,JMP,scalarfields2015JMP} -  which offers the possibility to explain all the assumptions 
of the {\it standard model}, with the appearance of families, the scalar higgs and the Yukawa
couplings included, and which offers also the explanation for the matter-antimatter asymmetry  in our 
universe and for the appearance of the dark matter -  a very simple  starting action for massless 
fermions and bosons in $d= (1+13)$ is assumed.
In this action
\begin{eqnarray}
{\cal A}\,  &=& \int \; d^dx \; E\;\frac{1}{2}\, (\bar{\psi} \, \gamma^a p_{0a} \psi) + h.c. +
%{\mathcal L}_{f} +  
\nonumber\\  
               & & \int \; d^dx \; E\; (\alpha \,R + \tilde{\alpha} \, \tilde{R})\,,%\nonumber\\
               %\end{eqnarray}
%
%\begin{eqnarray}
%{\mathcal L}_f &=& \frac{1}{2}\, (\bar{\psi} \, \gamma^a p_{0a} \psi) + h.c., 
%\nonumber\\
%p_{0a }        &=& f^{\alpha}{}_a p_{0\alpha} + \frac{1}{2E}\,
% \{ p_{\alpha}, E f^{\alpha}{}_a\}_-, 
%\nonumber\\  
%   p_{0\alpha} &=&  p_{\alpha}  - 
%                    \frac{1}{2}  S^{ab} \omega_{ab \alpha} - 
%                    \frac{1}{2}  \tilde{S}^{ab}   \tilde{\omega}_{ab \alpha},                   
%\nonumber\\ 
%R              &=&  \frac{1}{2} \, \{ f^{\alpha [ a} f^{\beta b ]} \;(\omega_{a b \alpha, \beta} 
%- \omega_{c a \alpha}\,\omega^{c}{}_{b \beta}) \} + h.c. \;, 
%\nonumber\\
%\tilde{R}      &=&  \frac{1}{2} \, \{ f^{\alpha [ a} f^{\beta b ]} \
%;(\tilde{\omega}_{a b \alpha,\beta} - 
%\tilde{\omega}_{c a \alpha} \,\tilde{\omega}^{c}{}_{b \beta})\} + h.c.\;, 
\label{wholeaction}
\end{eqnarray}
where $p_{0a } = f^{\alpha}{}_a p_{0\alpha} + \frac{1}{2E}\, \{ p_{\alpha},
E f^{\alpha}{}_a\}_- $, 
$ p_{0\alpha} =  p_{\alpha}  - \frac{1}{2}  S^{ab} \omega_{ab \alpha} - 
                    \frac{1}{2}  \tilde{S}^{ab}   \tilde{\omega}_{ab \alpha} $,                    
$R =  \frac{1}{2} \, \{ f^{\alpha [ a} f^{\beta b ]} \;(\omega_{a b \alpha, \beta} 
- \omega_{c a \alpha}\,\omega^{c}{}_{b \beta}) \} + h.c. $,  
$\tilde{R}  =  \frac{1}{2} \, \{ f^{\alpha [ a} f^{\beta b ]} \;(\tilde{\omega}_{a b \alpha,\beta} - 
\tilde{\omega}_{c a \alpha} \,\tilde{\omega}^{c}{}_{b \beta})\} + h.c.$, 
the two kinds of the Clifford algebra objects, $\gamma^a$ and $\tilde{\gamma}^a$,
\begin{eqnarray}
\label{twoclifford}
&&\{\gamma^a, \gamma^b\}_{+}= 2 \eta^{ab} = 
\{\tilde{\gamma}^a, \tilde{\gamma}^b\}_{+}\,,
\end{eqnarray}
which anticommute, $ \{\gamma^a, \tilde{\gamma}^b\}_{+}=0$ (they are connected with the 
left and the right multiplication of the Clifford objects, there is no third kind of operators ), are asumed, determining one 
of them spins and charges of spinors, another families.
Here~\footnote{$f^{\alpha}{}_{a}$ are inverted vielbeins to 
$e^{a}{}_{\alpha}$ with the properties $e^a{}_{\alpha} f^{\alpha}{\!}_b = \delta^a{\!}_b,\; 
e^a{\!}_{\alpha} f^{\beta}{\!}_a = \delta^{\beta}_{\alpha} $, $ E = \det(e^a{\!}_{\alpha}) $. 
Latin indices  
$a,b,..,m,n,..,s,t,..$ denote a tangent space (a flat index),
while Greek indices $\alpha, \beta,..,\mu, \nu,.. \sigma,\tau, ..$ denote an Einstein 
index (a curved index). Letters  from the beginning of both the alphabets
indicate a general index ($a,b,c,..$   and $\alpha, \beta, \gamma,.. $ ), 
from the middle of both the alphabets   
the observed dimensions $0,1,2,3$ ($m,n,..$ and $\mu,\nu,..$), indexes from 
the bottom of the alphabets
indicate the compactified dimensions ($s,t,..$ and $\sigma,\tau,..$). 
We assume the signature $\eta^{ab} =
diag\{1,-1,-1,\cdots,-1\}$.} 
$f^{\alpha [a} f^{\beta b]}= f^{\alpha a} f^{\beta b} - f^{\alpha b} f^{\beta a}$.
There are correspondingly for spinors two kinds of the
infinitesimal generators of the groups - $S^{ab}$ for $SO(13,1)$  and $\tilde{S}^{ab}$ for 
$\widetilde{SO}(13,1)$. 
 % are generators (App.~\ref{generatorsandfields}) 
%of the groups $SO(13,1)$ and $\widetilde{SO}(13,1)$, 
%
%\begin{eqnarray}
%\label{sabsc0}
The generators,
$S^{ab} = \,\frac{i}{4} (\gamma^a\, \gamma^b - \gamma^b\, \gamma^a)\,$ and 
$\tilde{S}^{ab} = \,\frac{i}{4} (\tilde{\gamma}^a\, \tilde{\gamma}^b - 
\tilde{\gamma}^b\, \tilde{\gamma}^a)$,
%\end{eqnarray}
% %expressible by $\gamma^a$ and $\tilde{\gamma}^a$, 
%
determine in the theory the spin and charges of fermions, $S^{ab}$, and the family quantum
numbers, $\tilde{S}^{ab}$.

The curvature $R$ and $\tilde{R}$ determine dynamics of gauge fields. 
% See Ref.~\cite{Norma2015}
% presented in 
%Sect.~\ref{lorentz} and in App.~\ref{auxilliaryappendix}. 

The infinitesimal generators of the Lorentz transformations for bosons operate as follows 
%
%\begin{eqnarray}
%\label{bosonspin0}
${\cal S}^{ab} \, A^{d\dots e \dots g} = i \,(\eta^{ae} \,A^{d\dots b \dots g} - 
\eta^{be}\,A^{d\dots a \dots g} )$. 
%\end{eqnarray}
%

We discuss in what follows properties of free massless fermion fields (Sect.~\ref{fermions}) and 
of free massless boson fields  (Sect.~\ref{bosons}), and suggest the interaction among fermions 
and bosons, which fulfill the Aratyn-Nielsen
theorem~\cite{holger}, but is in general not gauge invariance.

% Tell what will you do.
%06.11.2015 at 15:04

%
\subsection{Properties of general fermion fields}
\label{fermions}

 Let us make a choice of one kind of the Clifford algebra objects, let say $\gamma^{a}$'s, and
express correspondingly the linear vector space of fermions as follows
\begin{eqnarray}
\label{gammavector}
{\bf \Psi}(\gamma)&=& \psi + \sum_{k=1}^{d}\, \psi_{a_{1} a_{2} \dots a_{k}}\,
\gamma^{a_1}\gamma^{a_2} \dots \gamma^{a_k}\,,
\quad a_{i}\le a_{i+1}\,.
\end{eqnarray}

We could as well make a choice of $\tilde{\gamma}^a$'s instead of $\gamma^a$'s. 
We define  that operation of  $\gamma^a$ and $\tilde{\gamma}^a$ on such a vector space is 
understood as the {\it left} and the {\it right} multiplication, respectively, of any Clifford algebra 
object. Let 
 %(Eq.~(\ref{tildegcliffordappendix}))
$f (\gamma)$  be one of the (orthogonal) fermion states in the Hilbert space. The {\it left} and 
the {\it right} multiplication  can  be understood as follows
\begin{eqnarray}
\gamma^a \, f (\gamma) \,|\psi_{0} \,>: &=& (\, a_{0}\, \gamma^{a}\, + a_{a_1} 
\,\gamma^{a}\,\gamma^{a_1} + 
 a_{a_1 a_2}\,\gamma^{a}\, \gamma^{a_1} \gamma^{a_2} + a_{a_1 \cdots a_d}
 \,\gamma^{a}\,\gamma^{a_1} 
 \cdots \gamma^{a_d}\,)\,|\psi_{0} \,>\,, \nonumber\\
\tilde{\gamma}^a \, f (\gamma) \,|\psi_{0} \,>: &=& 
(\,i \,a_0 \gamma^a  -i\, a_{a_1} \gamma^{a_1}\, \gamma^a + i \,a_{a_1 a_2}
 \gamma^{a_1} \gamma^{a_2} \, \gamma^a 
+ \cdots + \nonumber\\
&&i\,(-1)^{d} \,a_{a_1 \cdots a_d} \gamma^{a_1} \cdots \gamma^{a_d}\, \gamma^a \,)\,
|\psi_{0}\,>\,,
\label{tildeclifford}
\end{eqnarray}
where $|\psi_{0}\,>$ is a vacuum state.

Eq.~(\ref{gammavector}) represents $2^d$ internal degrees of freedom, that is $2^d$ basic states.
Let us arrange the basis to be orthogonal in a way that operators $S^{ab}$ transform 
$2^{\frac{d}{2} -1}$ members of these basic states among themselves. They represent one family. 
The operators $\tilde{S}^{ab}$ transform each family member into the same family member of one 
of $2^{\frac{d}{2} -1}$ families. 

There are obviously four such groups 
of  $2^{\frac{d}{2} -1}$ families  with  $2^{\frac{d}{2} -1}$ family members ($2^{\frac{d}{2} -1}$
$\times$  $2^{\frac{d}{2} -1} $$\times 2^2=2^{d}$). These four groups differ in the eigenvalues
of the two operator of handedness, $\Gamma^{(1+(d-1))}$ and
$\widetilde{\Gamma}^{(1+(d-1))}$,  
\begin{eqnarray}
\label{GammatildeGamma}
\Gamma^{(1+(d-1))}= (-i)^{\frac{d-2}{2}} \, \gamma^{a_1} \gamma^{a_2}\dots 
\gamma^a_{d}\,,
\nonumber\\
\widetilde{\Gamma}^{(1+(d-1))}= (-i)^{\frac{d-2}{2}} \, \tilde{\gamma}^{a_1} 
\tilde{\gamma}^{a_2}
\dots \tilde{\gamma}^a_{d}\,,\nonumber\\
a_{k}< a_{k+1}\,.
\end{eqnarray}
The eigenvalues of [($\Gamma^{(1+(d-1))}$, $\widetilde{\Gamma}^{(1+(d-1))}$]  are 
$=[(+,+), (-,+), (+,-), (-,-)]$. Each of the groups can be 
extracted from the basis due to the requirements 
\begin{eqnarray}
\label{fourgroups}
A. \,\,\,
(1-\widetilde{\Gamma}^{(1+(d-1))})\, (1- \Gamma^{(1+(d-1))})\, {\bf \Psi} =0
%=(1- i^d\Gamma^{(1+(d-1))})\, {\bf \Psi}\, (1- \Gamma^{(1+(d-1))})
\,,\nonumber\\
B. \,\,\,
(1-\widetilde{\Gamma}^{(1+(d-1))})\, (1+ \Gamma^{(1+(d-1))})\, {\bf \Psi} =0
%=(1- i^d\Gamma^{(1+(d-1))})\, {\bf \Psi}\, (1+ \Gamma^{(1+(d-1))})
\,,\nonumber\\
C. \,\,\,
(1+\widetilde{\Gamma}^{(1+(d-1))})\, (1- \Gamma^{(1+(d-1))})\, {\bf \Psi} =0
%=(1+ i^d\Gamma^{(1+(d-1))})\, {\bf \Psi}\, (1- \Gamma^{(1+(d-1))})
\,,\nonumber\\
D. \,\,\,
(1+\widetilde{\Gamma}^{(1+(d-1))})\, (1+ \Gamma^{(1+(d-1))})\, {\bf \Psi} =0
%=(1+ i^d\Gamma^{(1+(d-1))})\, {\bf \Psi}\, (1+ \Gamma^{(1+(d-1))})
\,.%\nonumber\\
\end{eqnarray}
%
%The expressions on the right hands side of the equality sign follow from Eq.~(\ref{tildeclifford}), 
%due to the fact that $\widetilde{\Gamma}^{(1+(d-1))}$ and $ \Gamma^{(1+(d-1))}$ have in
%even dimensional spaces an even number of the Clifford algebra objects.
%07.11.2015 at 11.18

%CHECK again
In ($d=4n$)-dimensional spaces  the first and the last condition share the space of spinors 
determined by an even number of $\gamma^a$'s in each product,  Eq.~(\ref{gammavector}),
while the second and the third share the rest half of the spinor space determined by an odd 
number of $\gamma^a$'s in each product. In ($d=2(2n+1)$)-dimensional spaces is opposite:
The first and the last condition determine spinor space of and odd number of $\gamma^a$'s 
in each product, while the second and the third require an even number of $\gamma^a$'s 
in each product.

Let us denote these four groups of states, defined in Eqs.~(\ref{gammavector},\ref{fourgroups}) 
with the values of [($\Gamma^{(1+(d-1))}$, $\widetilde{\Gamma}^{(1+(d-1))}$]
$=[(+,+), (-,+), (+,-), (-,-)]$, by (${\bf \Psi}_{++}$, ${\bf \Psi}_{-+}$, 
${\bf \Psi}_{+-}$, ${\bf \Psi}_{--}$), respectively. 

States of each group can be chosen to fulfill the Weyl dynamical equation for free massless spinors
\begin{eqnarray}
\label{weyl}
\gamma^0\,\gamma^a p_{a} {\bf \Psi}_{ij}&=& 0\,, \nonumber\\
(i,j)& \in &\{ (+,+),  (-,+),  (+,-),  (-,-)\}\,.
\end{eqnarray}

 In the {\it spin-charge-family} theory one family contains, if analyzed with respect to the spin and 
charges of the {\it standard model}: the left handed weak charged quarks and leptons - electrons 
and  neutrinos - and the right handed weak chargeless quarks and leptons, with  by the {\it standard 
model} assumed hyper charges, as well as the right handed weak charged anti-quarks and 
anti-leptons and the 
left handed weak chargeless anti-quarks and anti-leptons. 

The break of the starting symmetry on both sectors, the spin and charges one and the families one,
than leads to two groups of four families, which gain masses at the electroweak break. 
All the rest families 
($2^{\frac{14}{2}-1}$ $ - 8 $) gain masses by interacting with the scalar fields.
%How can this go?

% How can I get $2^{\frac{13+1}{2}-1} - 8$ not being observed? How must the
%             break from $SO(13,1) $ to $SU(7,1) \times SU(3)\times U(1)_{II}$ be done that all but
%             $8$ families "disappear"? How $(4-1)\times (2^{\frac{14}{2}-1} \times
%              2^{\frac{14}{2}-1}2^{\frac{14}{2}-1})$ "disappear"? Are there only these eight
%             families which have the same $\Gamma^{(6)}$ and $\Gamma^{(6)}$?
%             How does this manifest in getting masses?

% $2^{6}$ members of one family representation of spinors - quarks and leptons 
%and antiquarks and antileptons, left and right handed, with spin in $d=(3+1)$ up and down - can be
%found in Table IV of Ref.~\cite{scalarfields2015JMP}. Each member of this representation can be 
%reached from a chosen one by $S^{ab}$. The eight families can be reached by $\tilde{S}^{ab}$,
%where $a $ and $b \in (0,1,\dots,8)$. In Subsect.~\ref{d2and4} the properties of basic states

%4.11.2015 at 18:23

These $2^d$ orthogonal basic states can be reached from any one of them 
%a choice of $f (\gamma)$ from Eq.~(\ref{tildeclifford}), 
by applying on such a state the products of
operators: a constant, $\gamma^{a_1}, \tilde{\gamma}^{a_1}$, and products of 
$\gamma^{a_i}$ and products of $  \tilde{\gamma}^{b_{1}}$. 

Let us see on  the case of $d=2$,  how do these four groups of families and family members
distinguish among  themselves. 

%Yet each of these four groups have particular properties. Let us discuss their properties for 
%$d=2$% and $d=4$ case
We shall check also conditions under which these fermion states fulfill the Weyl equation, 
(Eq.~(\ref{weyl})), for free (massless) fermions.

% 18.01.2016 at 4:12

%
\subsubsection{Properties of four groups of fermion states defined in Eq.~(\ref{fourgroups})}
\label{d2and4}

To better understand the meaning of the four groups (Eq.~(\ref{fourgroups})) of families and family 
members, let us start with the simplest case: $d=(1+1)$ - % and $d=(3+1)$ 
dimensional spaces. 

\vspace{3mm}

{\bf o} $\;\;\;$ {\bf  d=(1+1) case.} 

\vspace{3mm} 

The requirement A. of Eq.~(\ref{fourgroups})  ($(1-\tilde{\Gamma}^{(1+1)})\,
(1-\Gamma^{(1+1)})\, {\bf \Psi}=0$,\, ${\bf \Psi}_{++}= \psi + \gamma^0 \psi_0
+ \gamma^1 \psi_1 + \gamma^0 \gamma^1\, \psi_{01}$) leads to $\psi_0 + \psi_{1}=0 $, 
or consequently ${\bf \Psi}_{++}= \psi_{++}\, (\gamma^0 - \gamma^1)$. This state fulfills the
Weyl equation provided that $(p_0 -p_1)\, \psi_{++} =0$.

The requirement B. of Eq.~(\ref{fourgroups})  ($(1-\tilde{\Gamma}^{(1+1)})\,
(1+\Gamma^{(1+1)})\, {\bf \Psi}=0$) leads to $\psi + \psi_{01}=0 $, or
consequently ${\bf \Psi}_{+-}= \psi_{+-}\, (1-\gamma^0 \gamma^1)$. This state fulfills
the Weyl equation 
provided that $(p_0 +p_1)\, \psi_{+-} =0$.

The requirement C. of Eq.~(\ref{fourgroups})  ($(1+\tilde{\Gamma}^{(1+1)})\,
(1-\Gamma^{(1+1)})\, {\bf \Psi}=0$) leads to $\psi - \psi_{01}=0 $, or consequently
 ${\bf \Psi}_{-+}= \psi_{-+}\, (1+\gamma^0  \gamma^1)$. This state fulfills the Weyl
equation  provided that $(p_0 - p_1)\, \psi_{-+} =0$.

The requirement D. of Eq.~(\ref{fourgroups})  ($(1-\tilde{\Gamma}^{(1+1)})\,
(1-\Gamma^{(1+1)})\, {\bf \Psi}=0$) leads to $\psi_0 - \psi_{1}=0 $, or consequently
 ${\bf \Psi}_{--}= \psi_{--}\, (\gamma^0 + \gamma^1)$. This state fulfills the Weyl
equation  provided that $(p_0 +p_1)\, \psi_{--} =0$.

Making a choice of $p_1$ showing in the positive direction, the first and  the third choice 
correspond to the positive energy solution, while the second and the fourth choice correspond to the 
negative energy solution of the Weyl equation (\ref{weyl}).

% 13.11.2015 at 18:50
%The vacuum state has a negative energy. Should only the negative energy be counted?
%Then, since there are here only half of states negative, how is with boson states? Also half?
%What is with the negative norm of bosons discussed by Holger in the fermionization paper?
%
% Comment d=(13+1)

Each of the four groups of states contains $2^{\frac{d}{2}-1}=1$ state and $2^{\frac{d}{2}-1}
=1$ familiy. The operators $(1,\gamma^0 \gamma^1, \tilde{\gamma}^0 \tilde{\gamma}^1)$
are diagonal, the operators $(\gamma^0, \gamma^1, \tilde{\gamma}^0, \tilde{\gamma}^1) 
$ are off diagonal.
%transform the four states among themselves.
Let us present  the matrices for, let say, 
$\gamma^0$, $ \tilde{\gamma}^0$ and $\gamma^0 \tilde{\gamma}^0$ for the basic states,  
arranged as follows
$\stackrel{01}{(+ i)}= \frac{1}{2} (\gamma^0 -\gamma^1)$ (the case A.), 
$\stackrel{01}{(- i)}= \frac{1}{2}   (\gamma^0 +\gamma^1)$ (the case D.),
$\stackrel{01}{[+ i]}= \frac{1}{2} (1+ \gamma^0 \gamma^1)$ (the case C.),
$\stackrel{01}{[- i]} = \frac{1}{2} (1 -\gamma^0 \gamma^1)$ (the case B.). 
Let us notice that 
$\Gamma^{(1+1)}$ $(\stackrel{01}{(+ i)}, \stackrel{01}{(- i)},
\stackrel{01}{[+ i]}, \stackrel{01}{[- i]})=$ $(\stackrel{01}{(+ i)}, -\stackrel{01}{(- i)},
\stackrel{01}{[+ i]}, -\stackrel{01}{[- i]}) $, while   $\widetilde{\Gamma}^{(1+1)}$ 
$(\stackrel{01}{(+ i)}, \stackrel{01}{(- i)},\stackrel{01}{[+ i]}, \stackrel{01}{[- i]}) =$
$(\stackrel{01}{(+ i)}, -\stackrel{01}{(- i)},-\stackrel{01}{[+ i]}, \stackrel{01}{[- i]}) $. 
The notation arises from Refs.~\cite{hn02,hn03,DKhn} (See~\cite[Appendix]{scalarfields2015JMP}).
One finds the matrix representation for $\gamma^0$ and $\tilde{\gamma}^0$ and 
$\gamma^0 \tilde{\gamma}^0$
  \begin{equation}
  \label{gamma0}
   \gamma^{0} = \begin{pmatrix} 
 0 & 0 & 0 & 1\\
 0 & 0 & 1 & 0\\
 0 & 1 & 0 & 0\\
 1 & 0 & 0 & 0
  \end{pmatrix}\,,
 \tilde{\gamma}^{0} = \begin{pmatrix} 
 0 & 0 &  i & 0\\
 0 & 0 & 0 & i\\
 -i & 0 & 0 & 0\\
 0 &-i  & 0 & 0
  \end{pmatrix}\,,  
\gamma^0 \tilde{\gamma}^{0} = \begin{pmatrix} 
 0 & -i & 0 & 0\\
 -i & 0 & 0 & 0\\
 0 & 0 & 0 & i \\
 0 & 0 & i  & 0
  \end{pmatrix}\,.  
 \end{equation}

While $\gamma^0$ causes the transformations among states, which have the opposite handedness 
$\Gamma^{(1+1)}$, while they have the same  handedness $\tilde{\Gamma}^{(1+1)}$,
transforms $\tilde{\gamma}^0$  among states of opposite handedness $\tilde{\Gamma}^{(1+1)}$, 
leaving  handedness $\Gamma^{(1+1)}$ unchanged. The operator $\gamma^0 \tilde{\gamma}^0$
causes transformations among the states, which differ in both handedness. Interaction of the type
$S^{ab} \omega_{abc}$ and $\tilde{S}^{ab} \tilde{\omega}_{abc}$, appearing in the 
action,~Eq.(\ref{wholeaction}),  do not cause in this $d=(1+1)$ case transformations among the
basic states $(\stackrel{01}{(+ i)}, \stackrel{01}{(- i)}, \stackrel{01}{[+ i]}, \stackrel{01}{[- i]})$.

\vspace{3mm}

 {\bf o} $\;\;\;$ {\bf  d=(13+1) case.} 

\vspace{3mm}

In the case of $d=(13+1)$-dimensional space  the operators $S^{ab}$  transform all the members
of one family among themselves. Table~IV of Ref.~\cite{scalarfields2015JMP} represents one family
representation analyzed with respect to the {\it standard model} gauge and spinor groups. The
$2^{d/2-1} =64$ members represent quarks and leptons, left and right handed, with spin up and 
down and with the hyper charges as required by the {\it standard model}. There are also  the
anti-members, reachable from members not only by $S^{ab}$ but also by 
${\cal C}_{\cal N} {\cal P}_{\cal N}$~\cite{hnds}.

The operators $\tilde{S}^{ab}$ transform each family member of a particular family into another 
family, keeping the family member quantum numbers  unchanged.

There are four groups of such families, having  ($\Gamma^{(13+1)}$, $\tilde{\Gamma}^{(13+1)}$)
$=((+,+),$ $ (-,-),$ $ (+,-), (-,+))$, respectively.   As seen in the simple case of $d=(1+1)$ all four 
groups could be reachable from the starting one only by the operators $\gamma^a$, 
$\tilde{\gamma}^a$ and  $\gamma^a \tilde{\gamma}^b$.

We have some experience with the toy 
model in $d=(5+1)$, Refs.~\cite{DHN,DN012,familiesDNproc}, that when breaking symmetries not
only that only spinors of one handedness remain masless, but also most of families can get heavy 
masses.

After the break of $SO(13,1)$ to  $SO(7,1)$ $\times SO(6)$  (and correspondingly also of 
$\widetilde{SO}(13,1)$)  $S^{st}$, $s\in (0,\dots,8), t \in (9,\dots,14)$  (and correspondingly also 
of $\tilde{S}^{st}$, $s\in (0,\dots,8), t \in (9,\dots,14)$) are no longer applicable. 
% 14.11.2015 at 17:59 Four groups 
Anti spinors (spinors with quantum numbers of the second part, numerated by 33 up to 64, of 
Table IV in Ref.~\cite{scalarfields2015JMP}) are after the break reachable only by  
${\cal C}_{\cal N}$ ${\cal P}_{\cal N}$~\cite{hnds}. 

The break of $SO(6)$ to $SU(3)\times U(1)$ disables transformations from quarks to leptons. 
%15.11.2015 at 8:52  $\tilde{S}^{ab}$

When breaking symmetries, like from  $SO(13,1)$ to $SO(7,1)$ $\times SO(6)$, the break must
be done  in a way that only spinors of one handedness remain massless in order that the break
leads to observed (almost massless) fermions and that most of families get masses of the energy 
of the break~\cite{DHN,DN012,familiesDNproc}. Our studies so far support the assumption that only
the families with $\tilde{\Gamma}^{(7+1)}=1$ and $\tilde{\Gamma}^{(6)}=-1$ remain massless.

Correspondingly only eight families $(2^{(7+1)/2-1})$ remain massless.

At the further break of $SO(7,1)\times SU(3)\times U(1)$ to $SO(3,1)\times SU(2) $ 
$\times SU(3)\times U(1)$ all the eight families of quarks and leptons remain massless due  to 
the fact that left handed and right handed quarks and leptons have different charges and are 
correspondingly mass protected.

%Leave it out
\subsection{Properties of general boson fields}
\label{bosons}

We have discussed so far only fermion fields. The {\it spin-charge-family} theory action,
Eq.~(\ref{wholeaction}), introduces the vielbeins and the two kinds of the spin-connection fields
% - the gauge fields of the two kinds of the Clifford "charges", $S^{ab}$ and $\tilde{S}^{ab}$ - 
with which the fermions interact. These are the gauge fields of the two kinds of charges, which
take care  
of the family members quantum numbers $(S^{ab})$ - spins and charges - and of the family
quantum numbers $(\tilde{S}^{ab})$. 

The Lagrange density~(\ref{wholeaction}) of each kind of the spin connection fields is linear in 
the curvature. This very simple action is the extension of the action of the usual Kaluza-Klein kind 
of theories - having only one kind of the Clifford algebra operators. In this {\it spin-charge-family}
theory action  fermions carry the 
family and the family members quantum numbers, while the gravitational field - the vielbeins and the 
two kinds of the spin connection fields take care of the interaction among fermions. Vielbeins and 
spin connections are the only boson fields in the theory. They manifest at the low energy regime all
the phenomenologically needed vector and scalar bosons (explaining the higgs and the Yukawa
couplings).

Let us  define boson fields, which in the case of $d=(1+1)$,  $d=(13+1)$, or any $d$, transform 
the $2^{d}$  fermion states  among themselves. The fields $S^{ab} \omega_{abc}$ and 
$\tilde{S}^{ab} \tilde{\omega}_{abc}$ can, namely, cause transitions only among fermions with 
the same Clifford character: The Clifford even (odd) fermion states are transformed into the 
Clifford even (odd) fermion states, as we have seen in subsection~\ref{fermions}. 
%corresponding
%The  $2\times2^{d}$ boson fields, defined in Sect.~\ref{bosons},

Let us assume for this purpose that  there exist to each of products 
$\gamma^{a_1} \gamma^{a_2}\dots \gamma^{a_{k}} $, the number of products of 
$\gamma^{a}$'s running from zero to  $d$, the corresponding gauge fields: 
$\omega_{a_1 a_2 \dots a_k}$. There are obviously $2^d$ such gauge fields. These gauge 
fields, carrying $k$ vector indexes $a_{1} \dots a_{k}$, transform a fermion state 
${\bf \Psi}_{ij}$, $(i,j) =[(+,+), (-, +), (+, -), (-, -)] $ belonging to one of the four groups 
(with the eigenvalues of ($\Gamma^{(d)}$,$\tilde{\Gamma}^{(d)})$$=(i,j)$, respectively),  
discussed in subsection~\ref{fermions}, into another state, belonging to the same or to one of
the rest free groups:  If starting with the state of either the 
 A. or B. groups, these bosons transform  such a state to one of the states  belonging to either 
the group A. (if the number of $a_{j}$ is even) or to the group B. (if the number of $a_{j}$ is odd). 
If we start from the group C. or D., then the transformed state remains within these two groups.

Correspondingly we define to each of products $\tilde{\gamma}^{a_1}$ 
$ \tilde{\gamma}^{a_2}\dots \tilde{\gamma}^{a_{k}} $, again the number of products of 
$\tilde{\gamma}^{a}$'s running from zero to  $d$, the corresponding gauge fields 
$\tilde{\omega}_{a_1 a_2 \dots a_k}$, which again transform the state ${\bf \Psi}_{ij}$,  
belonging to one of the four groups, discussed in subsection~\ref{fermions}, into another state, 
belonging to the same  (if the number of $a_{k}$ is even),  or to one of the rest free groups 
 (if the number of $a_{k}$ is odd). In this case the transformations go from A. to C., or from 
B. to D..

All the states of one group of fermions are reachable from the starting state under the application
of $\omega_{abc}$ and $\tilde{\omega}_{abc}$. The operators $S^{ab} $ and $\tilde{S}^{ab}$
keep the handedness $\Gamma^{(d)}$ and $\tilde{\Gamma}^{(d)}$, respectively, unchanged. 
(Let us remind 
the reader that all the $2^{(13+1)/2 -1}$ states of one family (Table IV of 
Ref.~\cite{scalarfields2015JMP}) are reachable by $S^{ab} \omega_{abc}$ and all the
$2^{(7+1)/2 -1}$ families (Table V of Ref.~\cite{scalarfields2015JMP}) are reachable by 
$\tilde{S}^{ab} \tilde{\omega}_{abc}$).

% Check above paragraph. 20.01.11:45

The by the products of $\tilde{\gamma}^a$'s
transformed state ${\bf \Psi}$ differs in general from the one  transformed by the product of 
$\gamma^{a}$'s according to the definition in Eq.~(\ref{twoclifford}). 

Let us assume that all the boson fields obey the equations of motion
\begin{eqnarray}
\label{bosonsdeq}
\partial^a \partial_a \omega_{a_1 a_2 \dots a_k}&=&0\,,\nonumber\\
\partial^a \partial_a \tilde{\omega}_{a_1 a_2 \dots a_k}&=&0\,.
\end{eqnarray}
For the boson fields, which are the gauge fields of the products of $\tilde{\gamma}^{a_1}$ 
$ \tilde{\gamma}^{a_2}\dots \tilde{\gamma}^{a_{k}} $ or of 
$\gamma^{a_1} \gamma^{a_2}\dots \gamma^{a_{k'}} $ Eq.~(\ref{bosonsdeq}), this can only 
be true in the weak fields limit. 

Let us see the action of this boson fields on fermion basic states in the case of 
$d=(1+1)$.  The boson fields bring to fermions the quantum numbers, which they carry. We can
calculate these quantum numbers by taking into account Eq.~(16) in Ref.~\cite{scalarfields2015JMP} 
\begin{eqnarray}
\label{bosonspin0}
{\cal S}^{ab} \, A^{d\dots e \dots g} &=& i \,(\eta^{ae} \,A^{d\dots b \dots g} - 
\eta^{be}\,A^{d\dots a \dots g} )\,,
\end{eqnarray}
or we can simply calculate the action of the operators, the gauge fields of which are boson fields.
\begin{eqnarray}
\label{bosonsd2}
1 \;\;\, (1,\;\gamma^0, \gamma^1,\; \gamma^0\,\gamma^1)&=& 
(1,\gamma^0, \gamma^1, \gamma^0\,\gamma^1)\,,
\quad  \;\;\, \;\;\, \;\;\,
\tilde{1} \;\;\, (1,\gamma^0, \gamma^1, \gamma^0\,\gamma^1)= 
(1,\gamma^0, \gamma^1, \gamma^0\,\gamma^1)\,,  \nonumber\\
\gamma^0 \;\, (1,\gamma^0, \gamma^1, \gamma^0\,\gamma^1)&=& 
(\gamma^0, 1,  \gamma^0 \,\gamma^1,\gamma^1)\,,\quad \;\, \;\;\, \;\;\,
\tilde{\gamma}^0 \;\, (1,\gamma^0, \gamma^1, \gamma^0\,\gamma^1)= 
i\,(\gamma^0, -1, \gamma^0 \,\gamma^1,-\gamma^1)\,, \nonumber\\
\gamma^1 \;\, (1,\gamma^0, \gamma^1, \gamma^0\,\gamma^1)&=& 
(\gamma^1, - \gamma^0 \,\gamma^1,-1, \gamma^0)\,,\quad  \;\, 
\tilde{\gamma}^1 \;\, (1,\gamma^0, \gamma^1, \gamma^0\,\gamma^1)=
i\,(\gamma^1, - \gamma^0 \,\gamma^1, 1,- \gamma^0)\,, \nonumber\\
\gamma^0 \gamma^1\, (1,\gamma^0, \gamma^1, \gamma^0\,\gamma^1)
&=& (\gamma^0\,\gamma^1,- \gamma^1, -\gamma^0, 1)\, \quad 
\tilde{\gamma}^0\, \tilde{\gamma^1}\, (1,\gamma^0, \gamma^1, \gamma^0\,\gamma^1)
= (i)^2\,(\gamma^0\gamma^1, \gamma^1,\gamma^0,1)\,, 
\end{eqnarray}

{\it It is obvious that the two kinds of fields influence states in a different way, except the two 
constants, which leave  states untouched.} 

One can conclude that there are correspondingly $2\times 2^d -1$ independent real boson 
fields (only one of the two constants has the meaning), and there are also, as we have learned in 
Subsec.~\ref{fermions} $2^d$ complex fermion fields, which means $2\times 2^d$ real fermion 
fields in any dimension.  This supports the Aratyn-Nielsen theorem~\cite{holger}.

%
%\subsubsection{Comments for particular dimensions}
%
%\label{commentsparticulard}

\vspace{3mm}

{\bf o} $\;\;\;$ Comments on {\bf  d=(1+1) case}. 

\vspace{3mm}

Let us make a choice of $2^{\frac{d}{2}-1}$ fermion states, which is for $d=2$ only one 
state, say $\stackrel{01}{(+ i)}$.  It is the complex field and accordingly with two degrees of 
freedom.
One can make then (any) one choice of the boson field, let say $\omega_{01}$, which is the 
gauge field of the "charge" $\Gamma^{(1+1)} $. This is in agreement with the 
Aratyn-Nielsen theorem.

All the (complex) Clifford odd fermion states, ($\stackrel{01}{(+ i)}$, $\stackrel{01}{(- i)}$,
%$ \stackrel{01}{[+i]}$, $ \stackrel{01}{[-i]}$),
need three of the independent boson fields, let say 
(%$\omega, \gamma^0 \omega_{0}$,% 
$\gamma^1 \omega_{1}$, $\gamma^0  \gamma^1 \omega_{01}$,
% $\tilde{\gamma}^0 \tilde{\omega}_{0}$, 
$\tilde{\gamma}^1\tilde{\omega}_{1}$), 
%$\tilde{\gamma}^0 \tilde{\gamma}^1 \tilde{\omega}_{01}$ 
to be in agreement with the Aratyn-Nielsen theorem.

%All the (complex) fermion states, ($\stackrel{01}{(+ i)}$, $\stackrel{01}{(- i)}$,
%$ \stackrel{01}{[+i]}$, $ \stackrel{01}{[-i]}$), need all the independent boson fields, 
%($\omega, \gamma^0 \omega_{0}$, $\gamma^1 \omega_{1}$, 
%$\gamma^0  \gamma^1 \omega_{01}$, $\tilde{\gamma}^0 \tilde{\omega}_{0}$, 
%$\tilde{\gamma}^1\tilde{\omega}_{1}$, 
%$\tilde{\gamma}^0 \tilde{\gamma}^1 \tilde{\omega}_{01}$ to be in agreement with the
%Aratyn-Nielsen theorem.
%If we choose two states, let say $\stackrel{01}{(+ i)}, \stackrel{01}{[-i]}$, then three boson 
%fields must be chosen, like $\omega_{01}, \gamma^0 \omega_{0} and \gamma^1 \omega_{1}$. 

%
\subsubsection{Bosons in interaction with fermions}
\label{physicalbosons}
%

% Can it be that for some breaks of symmetries there are still the same numbres of bosons 
%and fermions?

If we expect gauge boson fileds to appear in the covariant derivative of fermions,  as we are used to
require, then all the gauge fields must curry the space index, like it is the case of  the 
covariant derivative for fermions, presented in Eq.~(\ref{wholeaction}): $p_{0a } = p_{a} - 
\frac{1}{2}  S^{bc} \omega_{bc a} - \frac{1}{2}  \tilde{S}^{bc}   \tilde{\omega}_{bc a} $.

Let us generalize this covariant momentum by replacing $\frac{1}{2}  S^{a_1 a_2} 
\omega_{a_1 a_2 a} + \frac{1}{2}  \tilde{S}^{a_1 a_2}   \tilde{\omega}_{a_1 a_2 a} $
by
\begin{eqnarray}
\label{covmom}
p_{0a}&=& p_a - \{
\omega_{a} + \gamma^{a_1} \omega_{a_1 a} + \gamma^{a_1} \gamma^{a_2} 
\omega_{a_1 a_2 a} + \dots +  \gamma^{a_1} \gamma^{a_2} \dots  \gamma^{a_d}  
\omega_{a_1 a_2 \dots a_d\, a} +\nonumber\\
&&\quad \quad\quad\quad \quad  \tilde{\gamma}^{a_1} \tilde{\omega}_{a_1 a} +
\tilde{\gamma}^{a_1} \tilde{\gamma}^{a_2}  \tilde{\omega}_{a_1 a_2 a} +
 \dots +  \tilde{\gamma}^{a_1} \tilde{\gamma}^{a_2} \dots \tilde{\gamma}^{a_d}  
\tilde{\omega}_{a_1 a_2\dots a_d\, a})\,. 
\end{eqnarray}
We assumed that all the $\gamma^a$'s in products appear in the ascending order. 
Correspondingly is $\frac{1}{2}  S^{a_1 a_2} \omega_{a_1 a_2 a} $ replaced by  $\frac{i}{2}$ 
$\gamma^{a_1} \gamma^{a_2}  \omega_{a_1 a_2 a}$, the factor $\frac{i}{2}$ appears due
to $S^{a_1 a_2}= \frac{i}{2} \gamma^{a_1} \gamma^{a_2}$, $a_2>a_1$.

This theory would neither be gauge invariant nor do the corresponding gauge fields fulfill the
equations of motion, Eq.~(\ref{bosonsdeq}), except in the weak limit if the gauge fields 
appear as the background fields. The degrees of freedom of bosons and fermions no longer 
fulfill the Aratyn-Nielsen theorem, 
unless we again allow either only Clifford even or Clifford odd fermion states and only 
one of the two fields with the space index zero, let say $\omega_{0}$ among the boson 
fields is allowed. 
And yet we have in addition nonphysical degrees of freedom due to gauge invariance for almost 
free massless fields in the weak limit, which should be possibly removed.

%There are now $2\times (d-2) \times 2^{d}-1$ degrees of freedom for real massless boson
%fields, $\omega_{a_1\dots a_k}$ and $\tilde{\omega}_{a_1\dots a_k}$ and $2\times 2^{d}$
%real massless fermion fields.
%
%The "vector" factor $d$ is replaced by $ (d-2)$  due to the required gauge invariance 
%$\partial^a \omega_{a_1 \dots a_{k}}=0 $ of the massless vector fields, fulfilling the equations
%of motion: $\partial^a \partial_a \omega_{a_1\dots a_k} =0$~\cite{mil} and the requirement
%that only the physical states are welcome. Equivalently for $\tilde{\omega}_{a_1\dots a_k}$. 

%In $d=(1+1)$ there are no degrees of freedom left and we are back with the previous case.

If nature has ever started with the boson fields as presented above, most of  these fields do not
manifest in $d=(3+1)$.

%In $d\ge 4$ we must first see whether all these fields are needed as physical fields. All the
%massless vector states, although having $(d=4)$ components, only two of them have physical
%meaning. (See what is already written about that. Can we have in this case massless scalar
%fields coupled to fermions? Not really....)

%In the {\it spin-charge-family} all the massless gauge fields have the vector index. They
%have additional (internal) quantum numbers. It is hard to see, whether some gauge fields with 
%the "scalar" index, that with the index of the higher dimensional part of space, have the same 
%properties for  $\omega$ and $\tilde{\omega}$ gauge fields, so that one of the two, having
%correspondingly no physical meaning, can be left out and whether it might be that with this 
%particular geuge field included the number of fermions and the the number of bosons would be
%equal.

%
\section{Conclusions}

We have started the fermionization of boson fields (or bosonization of fermion fields) in any $d$
(the reader can find the corresponding contribution in this proceedings) to understand better why,
 if at all, the nature has started in higher dimensions with the simple 
action as assumed in the {\it spin-charge-family} theory, offering in the low energy regime 
explanation for the appearance and the properties (what is in the {\it standard model} just 
assumed) of all the observed degrees of freedom of fermion and boson fields, with the families of 
the fermions, the scalar higgs and the Yukawa couplings included. This theory is a kind of the
Kaluza-Klein theories with two kinds of the spin connection fields. 
We also hope that the fermionization can help to see which role can the 
same number of degrees of freedom of fermions and bosons play in the
explanation, why the cosmological constant is so small. 

This contribution is a small step towards understanding better the open problems of the elementary 
particle physics and cosmology.  We discussed for any $d$-dimensional space the degrees 
of freedom for free massless fermions and the degrees of freedom for free massless bosons,
which are the gauge fields of all possible products 
of both kinds of the Clifford algebra objects, either of $\gamma^a$ or of $\tilde{\gamma}^a$. 
  
Although we have not yet learned enough to be able to answer the four questions, presented
in the introduction ({\bf a.}  Why is the simple starting action of the {\it spin-charge-family} theory 
doing so well in manifesting the observed properties of the fermion and boson fields? {\bf b.} Under
which condition can more general action lead to the starting action of Eq.~(\ref{wholeaction})?  
{\bf c.} What would more  general action, if leading to the same low energy physics, mean for the 
history of our Universe? {\bf d.} Could the fermionization procedure of boson fields or the bosonization procedure 
of fermion fields, discussed in this Proceedings for any dimension $d$ (by the authors of this 
contribution, while one of them, H.B.F.N.~\cite{holger},  has succeeded with another author to do
the fermionization for $d=(1+1)$),
% help to find the answers to the questions under  {\bf a. b. c.}?),
yet we have started to understand  the topic a little better.

\end{document}